\documentclass[11pt,a4paper,preprintnumbers,nofootinbib]{revtex4}

\pdfoutput=1

\usepackage[T1]{fontenc}
\usepackage[utf8]{inputenc}
\usepackage{amsmath}
\usepackage{amssymb}
\usepackage{booktabs}
\usepackage{hyperref}
\usepackage[final]{graphicx}
\usepackage{slashed}

\begin{document}

\title{The photon polarization in radiative $D$ decays, phenomenologically}
\author{Stefan de Boer$^{\,a}$}
\email{stefan.boer@kit.edu}
\author{Gudrun Hiller$^{\,b}$}
\email{ghiller@physik.uni-dortmund.de}
\affiliation{$^{\,a}$ Institut f\"ur Theoretische Teilchenphysik, Karlsruher Institut f\"ur  Technologie, D-76128 Karlsruhe, Germany\\
$^{\,b}$ Fakultät  Physik, TU Dortmund, Otto-Hahn-Str.4, D-44221 Dortmund, Germany}

\begin{abstract}
We work out the phenomenology of untagged time-dependent analysis with radiative $D^0$-decays into CP eigenstates, which allows to probe the photon polarization
by means of the charm mesons' finite width difference.
We show that  $D^0 \to \phi  \gamma$ or $D^0 \to \bar K^{0 *}  \gamma$ decays, which are SM-dominated, or SM-like, respectively, together with U-spin allow to 
obtain chirality-predictions for radiative decay amplitudes.  The order of magnitude  of wrong-chirality contributions in the SM  can be cross-checked with an up-down asymmetry in $D^0 \to \bar K_1^0 (\to \bar K \pi \pi) \gamma$.
We explore the sensitivity to new physics in   $|\Delta c|=|\Delta u|=1$  dipole couplings  in the decays $D^0 \to \rho^0 \gamma$. We point out the possibility to test the 
 SM with $D_s \to  K_1^+( \to K \pi \pi) \gamma$ decays.
\end{abstract}

\preprint{DO-TH 18/04, QFET-2018-04, TTP18-010}

\maketitle

\section{Introduction}

Rare charm decays provide a unique view to  flavor in the up sector, which, however, is mostly blurred by hadronic uncertainties.
These are particularly difficult to control in charm, as unlike in $K$- or $B$-physics, effective theory methods are not expected to work well.
Observables related to approximate symmetries of the standard model (SM), CP, lepton flavor conservation and universality are examples where nevertheless useful tests of the SM
can be performed. Here we investigate  the photon polarization in $|\Delta c|=|\Delta u|=1$ processes.
Short-distance contributions from the weak scale are expected to inherit the V-A-structure of the SM, a feature that  is generically not shared by SM extensions.
We propose to test the SM with the photon polarization in  $c \to u \gamma$ transitions.

Methods to extract the photon polarization can be inferred from $B$-physics  \cite{Atwood:1997zr,Melikhov:1998cd,Hiller:2001zj,Gronau:2001ng,Grossman:2000rk}.
These include the study of polarized $\Lambda_c$ hadrons in $\Lambda_c \to p \gamma$ decays  \cite{deBoer:2017que}, following the proposal for $\Lambda_b$'s \cite{Hiller:2001zj}.
Another possibility is to probe the photon dipole contribution in semileptonic decays  at very low dilepton invariant mass  with angular observables \cite{Melikhov:1998cd,Kruger:2005ep,Bobeth:2008ij,Becirevic:2011bp}.

In this work we study time-dependence in $D \to V \gamma$ decays, where $V$ denotes a vector meson, following a proposal for $B_s$-mesons  
\cite{Muheim:2008vu}~\footnote{We thank Jolanta Brodzicka for bringing this to our attention.}, and briefly discussed  in  \cite{Lyon:2012fk} for charm.
As first-principle theory predictions have large uncertainties,  we propose to use data and U-spin to 
 obtain  a data-driven SM prediction for the photon polarization in $D^0 \to V \gamma$, $V=\bar K^{*0}, \phi, \rho^0,\omega$. We work out the phenomenology, and provide predictions in models beyond the SM (BSM).
 We further suggest to study an up-down asymmetry in $D \to \bar K_1 ( \to \bar K \pi \pi) \gamma$  along the lines the one known to $B$-decays  \cite{Gronau:2001ng,Gronau:2002rz,Kou:2010kn,Gronau:2017kyq}, as a consistency check of the SM prediction for the photon polarization. An analogous asymmetry allows to 
 test the SM  with  $D_s \to  K_1^+ ( \to  K \pi \pi) \gamma$ decays.

The paper is organized as follows:
In section \ref{sec:time} we review time-dependence in  decays into CP-eigenstates and show how the photon polarization in $D \to V \gamma$  decays can be probed.
Features of different charm decay observables and their relations are discussed in section \ref{sec:anatomy}.
In section \ref{sec:strategy} we show how the SM can be tested and give BSM expectations.
In section \ref{sec:con} we summarize.  In the appendix we give  the angular distribution of $D_{(s)} \to  K_1 \gamma \to K  \pi \pi \gamma$ decays.

\section{Time-dependence in \texorpdfstring{$D \to V \gamma$}{DtoVgamma} \label{sec:time}}

The $D \to V \gamma $ decay  amplitudes can be written as
\begin{align}
 \mathcal A_{L,R}=\mathcal A(D\to V\gamma_{L,R})=\sum_j A_{L,R}^{(j)}e^{i\delta_{L,R}^{(j)}}e^{i\phi_{L,R}^{(j)}} \,,
\end{align}
where $L,R$ denote the chirality, $j$ labels different amplitudes, $A_{L,R}^{(j)} \geq 0$, $\delta_{L,R}^{(j)} $ are strong phases and $\phi_{L,R}^{(j)} $ are weak phases.
The corresponding CP-conjugated amplitudes are
\begin{align}
 & \bar{\mathcal A}_R=\text{CP}(\mathcal A_L)=\xi\sum_j A_L^{(j)}e^{i\delta_L^{(j)}}e^{-i\phi_L^{(j)}} \,, && \bar{\mathcal A}_L=\text{CP}(\mathcal A_R)=\xi\sum_j A_R^{(j)}e^{i\delta_R^{(j)}}e^{-i\phi_R^{(j)}} \,,
\end{align}
where $\xi$ denotes the CP eigenvalue of the self-conjugate vector meson $V$, {\it i.e.}~$\xi=+1$ for $V=\rho^0,\phi,\bar K^{*0}(K_S^0\pi^0)$ and $\xi=-1$ for $V=\bar K^{*0}(K_L^0\pi^0)$.

We define the normalized CP asymmetry as usual
\begin{align}
 \mathcal A_{\text{CP}}(D\to V\gamma)=\frac{\Gamma(D\to V\gamma)-\bar\Gamma(D\to V\gamma)}{\Gamma(D\to V\gamma)+\bar\Gamma(D\to V\gamma)} \, , 
\end{align}
where $\Gamma(D\to V\gamma)=\Gamma(D\to V\gamma_L)+\Gamma(D\to V\gamma_R)$.
The time-dependent decay rate is given as
\begin{align} \label{eq:TD}
 \Gamma(t)=\mathcal Ne^{-\Gamma t}\left(\cosh[\Delta\Gamma t/2]+A^\Delta\sinh[\Delta\Gamma t/2]+\zeta C\cos[\Delta mt]-\zeta S\sin[\Delta mt]\right) \,,
\end{align}
where $\zeta=+1$ for a $D$ meson, $\zeta=-1$ for a $\bar D$ meson and the normalization $\mathcal N$ can be found in, {\it e.g.}, \cite{Amhis:2016xyh}.
Here, $\Delta\Gamma=\Gamma_H-\Gamma_L>0$ and $\Delta m=m_H-m_L$ are the differences between the heavy and light $D$ mass eigenstates and $\Gamma$ is the mean width.
Note that different sign conventions and notations are used in the literature. 
The direct CP asymmetry $A_{CP}^{\rm dir}=C$ and the observable $S$ \cite{Muheim:2008vu} can be measured only when the initial flavor is tagged.
On the other hand, $A^\Delta$ can be observed in untagged time-dependent measurements by means of a finite width difference $\Delta \Gamma$, as has been shown already for the decays $B_s^0\to\phi\gamma$ \cite{Aaij:2016ofv}.

The observable $A^\Delta$ is given in terms of the decay amplitudes as
\begin{align}
 A^\Delta & =\frac{2\,\mathrm{Re}[\frac qp\left(\bar{\mathcal A}_L\mathcal A_L^*+\bar{\mathcal A}_R\mathcal A_R^*\right)]}{\left|\mathcal A_L\right|^2+\left|\mathcal A_R\right|^2+\left|\frac qp\right|^2\left(\left|\bar{\mathcal A}_L\right|^2+\left|\bar{\mathcal A}_R\right|^2\right)} \nonumber \\
 & = \frac1N4\,\xi\left|\frac qp\right|\sum_{j,k}A_R^{(j)}A_L^{(k)}\cos[\delta_R^{(j)}-\delta_L^{(k)}]\cos[\phi-\phi_R^{(j)}-\phi_L^{(k)}] \, , 
\end{align}
where
\begin{align}
 N & =\sum_{j,k}A_L^{(j)}A_L^{(k)}\left(\left(1+\left|\frac qp\right|^2\right)\cos[\delta_L^{(j)}-\delta_L^{(k)}]\cos[\phi_L^{(j)}-\phi_L^{(k)}]\right. \nonumber \\
 & \left.-\left(1-\left|\frac qp\right|^2\right)\sin[\delta_L^{(j)}-\delta_L^{(k)}]\sin[\phi_L^{(j)}-\phi_L^{(k)}]\right)+[L\leftrightarrow R] \,.
\end{align}
The 95\% C.L. intervals of the $D^0-\bar D^0$ mixing parameters read \cite{Amhis:2016xyh}
\begin{align} \label{eq:mixdata}
 & \left|\frac qp\right|\in[0.77,1.12] \,, && \phi =\text{Arg}(q/p)\in[-30.2,10.6]^{\circ}  \, , && \Delta\Gamma/(2\Gamma)\in[0.50,0.80]\% \,.
\end{align}

It  is instructive to consider $ A^\Delta $ in the limit $q/p\simeq 1$ and 
assuming that the decays can be described  by only one amplitude per chirality.
One obtains in this limit
\begin{align} \label{eq:one}
 A^\Delta & \simeq 2  \xi \frac{  A_L A_R }{ |A_L|^2 + |A_R|^2} \cos ( \delta_L-\delta_R)  \cos ( \phi_L -\phi_R) \, , 
\end{align}
where $A_a$, $\delta_a$ and $\phi_a$ denote the modulus, strong and weak phase, respectively of the chirality amplitude ${\cal{A}}_a=A_a e^{ i \delta_a} e^{ i \phi_a}$, $a=L,R$.
Eq.~(\ref{eq:one}) holds if there is no CP violation in the decay, or if  strong phases are negligible. As CKM-induced CP violation in charm is small  due to the GIM-mechanism this is a useful approximation within the SM
and in models with no  BSM sources of CP-violation.
Defining the photon polarization fraction $r$ as
\begin{align}
r = \frac{A_R}{A_L} \, ,
\end{align}
it follows
\begin{align}
 A^\Delta & \simeq 2  \xi \frac{  r }{ 1+r^2}   \cos ( \delta_L-\delta_R) \cos ( \phi_L -\phi_R)
\, . 
\end{align}
The polarization fraction in $D \to V \gamma $  decays can be extracted via $ A^\Delta$ obtained from the time-dependent distribution (\ref{eq:TD}) with an ${\cal{O}}(1 \%)$ coefficient (\ref{eq:mixdata}).
As direct CP violation requires the presence of both strong and weak phase,  a measurement of $A_{\rm CP}$ is complementary to $A^\Delta$.
In this work we consider only BSM models with negligible CP-violation.
The expression for $A^\Delta$   valid for this type of models including the full dependence  on the mixing parameters reads
\begin{align} \label{sec:master}
 A^\Delta & = \frac{4\,\xi\left|\frac qp\right|\cos\phi}{\left(1+\left|\frac qp\right|^2\right)}  \, \frac{ r }{ 1+r^2} \,  \cos ( \delta_L-\delta_R)
\, . 
\end{align}
We discuss expectations for the strong phases $\delta_{L,R}$  and relations between $D^0 \to V \gamma$ modes in section \ref{sec:anatomy}.

\section{Decay anatomies   \label{sec:anatomy}}

\begin{figure}
 \centering
 \includegraphics[trim=4cm 23.5cm 9.5cm 3.5cm,clip,width=0.8\textwidth]{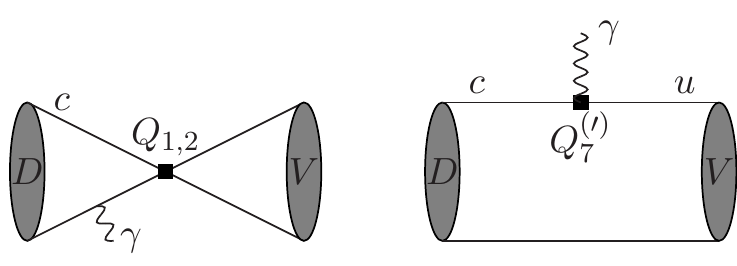}
 \caption{Weak annihilation (left) and short-distance (right) diagrams for $D\to V\gamma$ decays. There are additional diagrams (not shown) induced by $Q_{1,2}$ where the photon is emitted from other
 quark lines.}
 \label{fig:diagrams}
\end{figure}

The decays $D \to V \gamma, V=\bar K^{*0}, \phi, \rho^0,\omega$  are dominated in the SM by  weak annihilation (WA)  \cite{deBoer:2017que,Khodjamirian:1995uc,Lyon:2012fk}, see figure \ref{fig:diagrams}, plot to the left.
While this holds model-independently for $V=\bar K^{*0}$, the final state mesons $ \rho^0,\omega$ and, to a lesser degree,  the $\phi$ allow for additional contributions in and beyond the SM.
Here we consider BSM effects in dipole operators, 
\begin{align}
&Q_7=\frac{e\,m_c}{16\pi^2}(\bar u_L\sigma^{\mu_1\mu_2}c_R)F_{\mu_1\mu_2}\,,&&Q_7'=\frac{e\,m_c}{16\pi^2}(\bar u_R\sigma^{\mu_1\mu_2}c_L)F_{\mu_1\mu_2}\,
,\nonumber\\
 &Q_8=\frac{g_s\,m_c}{16\pi^2}(\bar u_L\sigma^{\mu_1\mu_2}T^ac_R)G^a_{\mu_1\mu_2}\,,&&Q_8'=\frac{g_s\,m_c}{16\pi^2}(\bar u_R\sigma^{\mu_1\mu_2}T^ac_L)G^a_{\mu_1\mu_2}\,,
\end{align}
in the effective Lagrangian
\begin{align}
 \mathcal L_\text{eff}^\text{weak}=\frac{4G_F}{\sqrt 2}\left(\sum_{q=d,s}V_{cq}^*V_{uq}\sum_{i=1}^2C_iQ_i^{(q)}
 +\sum_{i=7}^8\left(C_iQ_i+C_i'Q_i'\right)\right)\,,
\end{align}
where $G_F$ is the Fermi constant and $V_{ij}$ are CKM matrix elements.
The left- and right-handed Wilson coefficients $C_{7,8}$ and  $C_{7,8}^\prime$, respectively, are purely BSM as their SM contributions vanish by  GIM-cancellations.
The chromomagnetic operators $Q_8^{(\prime)}$ enter radiative decay amplitudes at higher order \cite{deBoer:2017que,Dimou:2012un,Lyon:2012fk}, but there is a  contribution from  mixing
onto  $Q_7^{(\prime)}$. It can be absorbed effectively  into the coefficient  $C_7^{(\prime)}$, see, {\it e.g.}, \cite{deBoer:2017que} for explicit  formulae.
Corresponding contributions to $D \to V \gamma$ are illustrated in figure \ref{fig:diagrams},  plot to the right.
The four-fermion operators $Q_{1,2}^{(q)} \sim  \bar u_L \gamma_\mu q_L \bar q_L  \gamma^\mu c_L$ are SM-like and induce WA amplitudes. It is possible that chirality-flipped versions of $Q_{1,2}^{(q)}$ are present in BSM scenarios.
As we neglect CP-violation such contributions are not distinguishable  from the V-A ones, and effectively accounted for in our framework.

To test the SM using $ A^\Delta$ requires sufficient understanding of its SM value -- it is the main point of this paper to obtain such a prediction experimentally by relating
$ A^\Delta$ from SM-dominated modes $V=\bar K^{*0}, \phi$ to $ A^\Delta$ from $V= \rho^0,\omega$ using U-spin.
We show in section  \ref{sec:strategy} that this framework describes available data in a consistent way.

In the following we give details on  $D^0 \to V \gamma$ decays for  $V= \bar K^{*0}, \rho^0$ and $\phi$  in section \ref{sec:barK}, 
\ref{sec:rho0} and  \ref{sec:phi}, respectively. These decays
enter the  SM  tests described in section \ref{sec:strategy}.
In section \ref{sec:D0}  and  \ref{sec:Ds} we discuss  $D^0 \to \bar K_1^0( \to \bar K \pi \pi) \gamma$  and $D_s \to  K_1^+( \to  K \pi \pi) \gamma$ decays, respectively.
The former mode assists the extraction of the SM's photon polarization from $ A^\Delta$ as argued in section \ref{sec:barK} as well as serves as a standard candle for
BSM searches with the latter, the $D_s$-decays.

\subsection{\texorpdfstring{$D^0\to\bar K^{*0}\gamma$}{D0tobarKstar0gamma} decays}\label{sec:barK}

The decay $D^0\to\bar K^{*0}\gamma$ is purely  induced by WA and SM-like. Strong phases are small
 \cite{deBoer:2017que,Khodjamirian:1995uc,Lyon:2012fk},
however, beyond leading order effects  could induce non-vanishing phases~\footnote{ Electromagnetic (soft) final state phases can be neglected \cite{Atwood:1997zr}.}.
Note that  predictions for the  $D^0\to\bar K^{*0}\gamma$  branching ratio  obtained in  other, hybrid chiral
 frameworks \cite{Burdman:1995te,Fajfer:1997bh,Fajfer:1998dv} are in line with experimental data given in  table \ref{tab:br} only if interfering amplitudes add coherently  \cite{deBoer:2017que},
 {\it  i.e.}, for small relative strong phases.
Therefore, eq.~(\ref{sec:master}) simplifies 
\begin{align}
 A^\Delta(D^0\to\bar K^{*0}\gamma) \simeq \frac{4\,\xi_{\bar K^{*0}}  \left|\frac qp\right|\cos\phi}{\left(1+\left|\frac qp\right|^2\right)}  \frac{r_0}{1+r_0^2} \,,
\end{align}
where $r_0$ denotes the corresponding $D^0\to\bar K^{*0}\gamma$ photon polarization fraction. Theoretical predictions for $r_0$ are rather uncertain as 
$A_R$ is presently not known, except for being  power suppressed with respect to $A_L$ \cite{deBoer:2017que},
\begin{align}
r_0=O\left(\frac{\Lambda_{\text{QCD}}}{m_c}\right) \, . 
\end{align}
Due to the low charm mass corrections can be considerable and $r_0$ unsuppressed \cite{Lyon:2012fk}.
As $r_0<1$ for a convergent power series a measurement of $r_0$ allows to probe the performance of the expansion.

In view of the sizeable uncertainties we refrain from using theory input on $r_0$ and propose to use the value that can be determined experimentally via $A^\Delta$.
To do so we  assume that the strong phase {\it difference} $\delta _L-\delta_R$  and $r_0$ are not both large,  because in this case a suppressed $A^\Delta$ could not
unambiguously point to a suppressed $r_0$, see eq.~(\ref{sec:master}).
This possibility, although being not the plain-vanilla theory expectation, can only be cross-checked  with other measurements:

We propose to study an up-down asymmetry in $D^0 \to \bar K_1 \gamma$ with $\bar K_1 \to \bar K \pi \pi$, constructed from the photon with respect to the plane formed by the pions in the $\bar K_1$'s rest frame, discussed  further in section \ref{sec:D0}, and defined in the appendix.
The advantage of measuring the up-down asymmetry in $D^0 \to \bar K_1 \gamma$ decays  is that it probes the photon polarization parameter
\begin{align} \label{eq:asy}
\lambda_\gamma=- \frac{1-r_0^2(\bar K_1)}{1+r_0^2(\bar K_1)} \, , 
\end{align}
which is independent of the relative phase between $A_L$ and $A_R$.
On the flip-side,  $D^0$-tagging is required, because $\lambda_\gamma$  changes sign between $D^0$ and $\bar D^0$. Note, $K^- \pi^+ \pi^0$ final states  are self-tagging, unlike $ \bar K^0 \pi^+ \pi^-$ ones. This method returns predominantly the polarization fraction of the $\bar K_1(1270)$-resonance, $r_0(\bar K_1)$, rather than  of the one of the  vector $\bar K^{*0}(892)$, $r_0$. As spin, flavor and color counting are identical  and masses not too much apart we expect the dynamics to  be sufficiently related. 
There is a doubly-Cabibbo suppressed contamination from $D^0 \to  K_1^0 \gamma$, $ K_1 \to  K \pi \pi$ decays affecting $\lambda_\gamma$
at order $V_{cd}^* V_{us}$, that is, a few percent.
One may also  use the doubly-Cabibbo suppressed but color-enhanced modes  $D^+ \to  K_1^+ \gamma$  to estimate the size of the SM polarization.

Another way is to look for large relative (strong) phases with CP-asymmetries
in the $D^0\to\rho^0\gamma$ time-dependent distribution (\ref{eq:TD}), which
are sensitive to phases in a complementary way. The last method requires to establish  a finite  CP-asymmetry in $D^0 \to \rho^0 \gamma$.
The current measurement   $A_{\rm CP}(D^0 \to \rho^0 \gamma)=0.056\pm0.152\pm0.006$ \cite{Abdesselam:2016yvr} is consistent with zero.
We note that the phase differences probed in CP-asymmetries are those corresponding to  {\it same} chirality amplitudes, so the relation to $\delta_L-\delta_R$ is not
immediate.

\subsection{\texorpdfstring{$D^0\to\rho^0\gamma$}{D0torho0gamma} decays}\label{sec:rho0}

The WA-contributions of  $D^0\to\rho^0\gamma$ and $D^0\to\bar K^{*0}\gamma$  are related by U-spin.
Therefore, in the SM,
\begin{align} \label{eq:Uspinanatomy}
A_{L,R}^{\text{SM}} (\rho^0)=A_{L,R} (\bar K^{*0})  \times \text{[U-spin corrections]} \, . 
\end{align}
Here we neglected contributions from the soft gluon operator $c \to u \gamma g$ \cite{Grinstein:2004uu}, see also \cite{Becirevic:2012dx},  to $D \to \rho^0 \gamma$, where it is GIM-suppressed \cite{deBoer:2017que}.
The perturbative and hard spectator interaction induced SM-amplitudes for $c\to u$ transitions are negligible with respect to the WA-amplitude \cite{deBoer:2017que}.

While the U-spin breaking from differences in masses and CKM elements can be accounted for trivially, the residual one on the left and right-chiral amplitude, denoted by $f_L$, $f_R$, respectively, depends on hadronic physics. Note, $f_{L,R}$ are in general complex-valued.
Estimations based on factorization identify  the largest WA-contributions as the ones with the photon  being radiated off the initial state  
\cite{Bosch:2001gv,deBoer:2017que,Khodjamirian:1995uc}.
In this case, the breaking in the matrix element is given by the final vector meson's matrix element, $\langle V |\bar q \gamma_\mu q^\prime|0 \rangle \propto
m_V f_V$.   For the modes at hand, $f_{L,R}=m_\rho f_\rho/(m_{K^{*0}} f_{K^*}) \simeq 0.9$, 
an effect within the  nominal  size of  
U-spin breaking  in charm, ${\cal{O}}(0.2-0.3)$, {\it e.g.}, \cite{Brod:2012ud,Hiller:2012xm,Muller:2015lua}.
We find that
in the hybrid model \cite{Fajfer:1997bh,Fajfer:1998dv}, also \cite{Casalbuoni:1992dx}, using the expressions compiled in  \cite{deBoer:2017que},  the U-spin breaking is  of similar size,
$f_{L,R}\simeq0.9 \pm 0.1$, where we varied input parameters.

From  (\ref{eq:Uspinanatomy}) follows
 \begin{align} \label{eq:SM}
 r^{\rm SM}=r_0   \, ,
 \end{align}
 subject to corrections of the order $f_R/f_L$. Eq.~(\ref{eq:SM})
provides, once  $r_0$ is known from $D^0\to\bar K^{*0}\gamma$ data, a SM-prediction for $D^0 \to \rho^0 \gamma$.
Hence, up to U-spin breaking,
\begin{align}\label{eq:ADelta_equality}
 A_{\text{SM}}^\Delta(D^0\to\rho^0\gamma)   \simeq    \xi_{\bar K^{*0}}  \xi_{\rho^0}    A^\Delta(D^0\to\bar K^{*0}\gamma) \,.
\end{align}
Any sizeable deviation from eq.~(\ref{eq:ADelta_equality}) would signal  BSM physics in the $c\to u$ transition which contributes to $D^0\to\rho^0\gamma$, but not to $D^0\to\bar K^{*0}\gamma$.~
On the other hand, experimental confirmation of  eq.~(\ref{eq:ADelta_equality}) would establish  $c\to u\gamma$ amplitudes  other than  WA  ones to  be subleading.

\subsection{\texorpdfstring{$D^0\to\phi\gamma$}{D0tophigamma} decays  \label{sec:phi}}

The decay $D^0\to\phi\gamma$ is not a pure WA-induced decay due to the $d \bar d +u\bar u$ admixture, or rescattering  \cite{Isidori:2012yx}.
We parameterize such effects by a complex-valued parameter $y$, and $y \lesssim O(0.1)$  as follows \cite{deBoer:2017que} 
\begin{align}
 A_{L,R} (\phi)  \simeq  A_{L,R}^{\rm WA} (\phi) + y \left(      A_{L,R}^{\rm WA} (\rho^0) -   A_{L,R}^{\rm 7,8} (\rho^0)    \right) \, , 
\end{align}
where $A_{L,R}(\rho^0) = A_{L,R}^{\rm SM}{(\rho^0)} +  A_{L,R}^{\rm 7,8} (\rho^0) $. Here $ A_{L,R}^{\rm 7,8}$ denote contributions from dipole operators $Q^{(\prime)}_{7,8}$. The  different sign  between the $\phi$ and the $\rho^0$  arises from the  $SU(3)$-decomposition.
Up to U-spin breaking between the $\phi$ and the $\bar K^{*0}$ of the order $f_R/f_L$ 
holds
\begin{align} \label{eq:rfi}
r_\phi=r_0 (1+{\cal{O}}(y)) \, ,
\end{align}
where $r_\phi$ denotes the polarization fraction of the photon in $D \to \phi \gamma$ decays. Therefore,
\begin{align}
 A^\Delta(D^0\to\phi\gamma)\simeq\xi_{\bar K^{*0}}  \xi_\phi A^\Delta(D^0\to\bar K^{*0}\gamma) \, (1 +{\cal{O}}(y) ) \,  .
\end{align}
As  already discussed in section \ref{sec:rho0} for the $\rho^0$, the 
leading U-spin breaking based on dominance of initial state radiation is given by $f_{L,R} =m_\phi f_\phi/(m_{K^{*0}} f_{K^*}) \simeq 1.3$. Similarly,  the numerical agreement with 
 the  hybrid model is good. We find $f_{L,R}\simeq1.2 \pm 0.1$, where the amplitudes have been added coherently.
 For destructive interference, which is in conflict  with data on the $D^0 \to \phi \gamma$ branching ratio, $f_{L,R}\simeq1.5$.

When $y$ can be neglected with respect to other uncertainties or contributions,  $D^0\to\phi\gamma$ becomes a standard candle very much like  $D^0\to \bar K^{*0}  \gamma$.
At higher precision, at ${\cal{O}}(y)$, the decay $D^0\to\phi\gamma$  becomes  sensitive to BSM physics similar to $D^0\to\rho^0\gamma$.

\subsection{\texorpdfstring{$D^0 \to \bar  K_1^0 \gamma$}{D0tobarK1gamma} decays \label{sec:D0}}

The up-down asymmetry in $D^0 \to \bar K_1^0 \gamma$ with $\bar K_1^0 \to \bar K \pi \pi$ probes the photon polarization by measuring the  polarization
of the kaon resonance $\bar K_1^0$. 
The asymmetry is the $D$-decay version of the one in
$B \to K_1 \gamma$ decays \cite{Gronau:2002rz}. 
It is proportional to the photon polarization parameter $\lambda_\gamma$ (\ref{eq:asy}), see the appendix for details.
The  proportionality factor depends on the details of hadronic decays of the $\bar K_1$.
As such, it is independent of the resonance's production, hence, is the same for $B$- and $D$-decays. The rate of $\bar K \pi \pi$ events, of course,  differs as well as the relative importance of
resonances and their interference effects.

The  contribution from the $K(1400$)-family and higher, which includes  resonances which dilute the asymmetry,  is phase space suppressed   in charm  relative to the  $\bar K_1(1270)$ one  by  about a factor of two.
This reduces the impact of interference effects and suggests a single-resonance analysis in terms of the $\bar K_1(1270)$.
As stressed in the more recent $B$-physics literature \cite{Kou:2010kn,Aaij:2014wgo,Kou:2016iau,Gronau:2017kyq}, 
insufficient understanding of the hadronic structure of the $\bar K_1$-decay prohibits a precision
extraction of the photon polarization. While this can be overcome \cite{Kou:2016iau,Gronau:2017kyq}, here, we merely need to check whether  the wrong-chirality amplitudes satisfy $A_R \sim A_L$ or not in a SM-like decay. 

The proportionality factor between the integrated up-down asymmetry (\ref{eq:updown}) and $\lambda_\gamma$ has been estimated for the $\bar K_1(1270) \to \bar K^0 \pi^+ \pi^-$ to be within $-13\%$ to $+24 \%$,
and for  $\bar K_1(1270) \to K^-  \pi^+ \pi^0$  to be around $-(7-10) \%$  \cite{Gronau:2017kyq}. Measurement of a near maximal asymmetry would imply a small $r_0(\bar K_1)$.
 A detailed analysis of $\bar K_1$-distributions  is beyond the scope of this work.

\subsection{\texorpdfstring{$D_s \to  K_1^+ \gamma$}{DstoK1gamma} decays \label{sec:Ds}}

The decay $D_s \to K_1^+ ( \to K \pi \pi) \gamma$  is  color-allowed, hence the sensitivity to BSM  physics is suppressed by $1/N_C$, where $N_C$ denotes the number of colors, relative to the one in
$D^0 \to ( \rho^0, \omega) \gamma$ decays. A similar up-down asymmetry as in $D^0 \to \bar K_1^0 ( \to \bar K \pi \pi) \gamma$ can be constructed, see appendix.
Predictions for the polarization fraction of $D_s \to   K_1^+ \gamma$ decays, $r_s( K_1)$, are
\begin{align}
r_s( K_1) =r_0(\bar K_1) (1 +{\cal{O}}(1/(V_{us} N_C)) ~~~~\mbox{or}~~  r_s( K_1)\sim {\cal{O}}(1/N_C) r ~~~~\mbox{if} ~~ r_0 \simeq 0 \, .
\end{align}
In the latter case, for negligible SM-contribution to $A_R$,  $\lambda_{\gamma s} +1$ becomes a null test of the SM, where $\lambda_{\gamma s}$ denotes the corresponding  photon polarization parameter (\ref{eq:asy}) 
in $D_s \to   K_1^+ \gamma$ decays
\begin{align} \label{eq:asys}
\lambda_{\gamma s}=- \frac{1-r_s^2( K_1)}{1+r_s^2( K_1)} \, .
\end{align}
In the SM it has to be equal to $\lambda_\gamma$, eq.~(\ref{eq:asy}),  up to U-spin corrections.
Significant deviations can signal BSM physics.

\section{Testing the  SM \label{sec:strategy}}

We  provide explicit expressions  on how to probe  BSM-sensitive contributions in the photon polarization fraction in $D^0 \to \rho^0 \gamma$ decays using $D^0 \to \bar K^{*0} \gamma$ in section \ref{sec:Kstar}, and using $D^0 \to \phi \gamma$  in section \ref{sec:fi}. We also show consistency of the framework -- WA dominance in SM-dominated modes consistent  with  leading U-spin breaking
 -- with current data on branching ratios, and give expectations
for the photon polarization in  $D \to \rho^0 \gamma$ decays in BSM models.

\subsection{\texorpdfstring{$D^0 \to \bar K^{*0}  \gamma$}{D0tobarKstar0gamma} \label{sec:Kstar}}

The $D^0 \to \bar K^{*0}  \gamma$  is a  WA-induced mode. Its  branching ratio can be written as 
\begin{align}
{\cal{B}}(D^0 \to \bar K^{*0}  \gamma)  & = \tau_{D^0} \frac{m_{D_0}^3}{32 \pi} \left(1-\frac{m_{K^*}^2}{m_{D_0}^2} \right)^3  \frac{ \alpha_e (G_F m_c )^2}{\pi^3}   \cdot {\cal{B}}_0 \, ,  \\
{\cal{B}}_0& =  |a_0|^2 +|a_0^{\prime}|^2  \,, \label{eq:B}
\end{align}
where, in the notation of the previous sections, $a_0,a_0^\prime$ correspond to $A_L,A_R$, respectively.
Then, the ratio of right-  to left-handed photons is given as
\begin{align}
r_0 = \left|\frac{a_0^{\prime}}{a_0}\right| \, .
\end{align}
A measurement of ${\cal{B}}_0, r_0$ returns the magnitude of both amplitudes
\begin{align}
&|a_0| =\sqrt{ \frac{  {\cal{B}}_0   }{1+r_0^2}} \, , \qquad  |a_0^{\prime}| =r_0 \sqrt{ \frac{  {\cal{B}}_0   }{1+r_0^2}} \, . 
\end{align}

The BSM-sensitive mode $D^0 \to \rho^0  \gamma$  can be affected by contributions from left- and right-handed Wilson coefficients $C_7$ and  $C_7^\prime$, respectively.
We write the branching ratio as (note, factor $1/2$ for isospin)
\begin{align} \nonumber
{\cal{B}}(D^0 \to \rho^0 \gamma)  &  = 1/2 \,  \tau_{D^0} \frac{m_{D_0}^3}{32 \pi} \left(1-\frac{m_{\rho}^2}{m_{D_0}^2} \right)^3  \frac{ \alpha_e (G_F m_c )^2}{\pi^3} \cdot {\cal{B}} \, ,  \\
{\cal{B}} & =  |a + T C_7|^2 + |a^{\prime} + T C_7^\prime |^2  \, ,  \label{eq:rhoB}\\
a^{ (\prime)} & =   - \frac{V^*_{cd}}{V^*_{cs}}  a_0^{ (\prime)} \times f_{L (R)} \label{eq:Uspin} \, ,
\end{align}
where $T$ denotes the $D^0 \to \rho^0$ dipole form factor at maximum momentum transfer, and $f_{L,R} \neq 1 $  accounts for U-spin breaking effects beyond phase space and CKM
already discussed in section \ref{sec:anatomy}.
The polarization fraction of $D^0 \to \rho^0 \gamma$ is given as 
\begin{align}
r  = \left|\frac{a^{\prime}+T C_7^\prime}{a+T C_7}\right| \, .
\end{align}
Experimental findings for the reduced branching ratios ${\cal{ B}}_0$, ${\cal{ B}}$ and ${\cal{ B}}_\phi$, the latter corresponding to $D \to \phi \gamma$ decays,  are given in  table \ref{tab:br}.

Measurement of  4 observables, ${\cal {B}}, {\cal{B}}_0,r,r_0$ determines 4 coefficients, the SM contributions $a, a^{ \prime}$ and the BSM ones $C_7, C_7^\prime$.
By definition, $r,r_0 \geq 0$.
Presently, only  branching ratios are measured, see table \ref{tab:br}. It would be desirable to have more precise data available, in particular, the discrepancy  in 
$D^0\to\bar K^{*0}\gamma$ between Belle and BaBar should be settled. 

\begin{table}
 \centering
 \begin{tabular}{ccccc}
  \toprule
  branching ratio                                                                                        &  $D^0\to\rho^0\gamma$          &  $D^0\to\omega\gamma$  &   $D^0\to\phi\gamma$           &  $D^0\to\bar K^{*0}\gamma$  \\
  \midrule
  Belle \cite{Abdesselam:2016yvr}$^\dagger$                                                              &  $(1.77\pm0.31)\times10^{-5}$  &  --                    &  $(2.76\pm0.21)\times10^{-5}$  &  $(4.66\pm0.30)\times10^{-4}$  \\
  BaBar \cite{Aubert:2008ai}$^\dagger$\footnote{We update the normalization \cite{Patrignani:2016xqp}.}  &  --                            &  --                    &  $(2.81\pm0.41)\times10^{-5}$  &  $(3.31\pm0.34)\times10^{-4}$  \\
  CLEO \cite{Asner:1998mv}                                                                               &  --                            &  $<2.4\times10^{-4}$   &  --                            &  --  \\
  \midrule
  $\cal{B}^{\text{Belle}}$                                                                               &  $0.030\pm0.005$               &  --                    &  $0.039\pm0.003$               &  $0.49\pm0.03$  \\
  $\cal{B}^{\text{BaBar}}$                                                                               &  --                            &  --                    &  $0.039\pm0.006$               &  $0.35\pm0.04$  \\
   \bottomrule
 \end{tabular}
 \caption{Experimental data on $D^0\to V\gamma$ branching ratios.
 The corresponding numerical values for the reduced branching ratios $\cal{B}$, see eqs.~(\ref{eq:B},\ref{eq:rhoB}) and analogously  for $\phi \gamma$, are given in the last row.
 $^\dagger$Statistical and systematic uncertainties are added in quadrature.}
 \label{tab:br}
\end{table}

In absence and anticipation of future polarization data we discuss the following limiting cases:

\begin{itemize}
\item[a)] $C_7, C_7^\prime \simeq 0$.  This corresponds to the SM, $r \simeq r_0$, discussed around eq.~(\ref{eq:SM}).
\item[b)] $r_0 \simeq 0$. It follows
\begin{align} \label{eq:r0zero}
r=\frac{|T C_7^\prime|}{ \sqrt{ {\cal{B}}-|T C_7^\prime|^2}} \, . 
\end{align}
The polarization fraction $r$ is a null test of the SM for negligible $r_0$.
We can already now make a  data-based prediction for $r$ given $C_7^\prime$ irrespective of $C_7$.  Possible values of $r$ from eq.~(\ref{eq:r0zero}) are illustrated in figure \ref{fig:r0zero}, where the blue band displays  the one sigma
range of ${\cal{B}}$.
Within leptoquark models holds $|C_7'|\lesssim0.02$, which, using $T=0.7$ \cite{deBoer:2017que},
 implies $r\lesssim0.09$, indicated by the green box.  On the other hand, SUSY models can 
provide significantly higher values $|C_7'|\lesssim0.3$, while model-independently holds $|C_7'|\lesssim0.5$. 
As $r$ diverges towards $C_7^\prime\simeq0.15$, in both latter cases there is no upper limit on $r$.
Upper limits  on the Wilson coefficients are taken from \cite{deBoer:2017que}.  
\begin{figure}
 \includegraphics[width=0.45\textwidth]{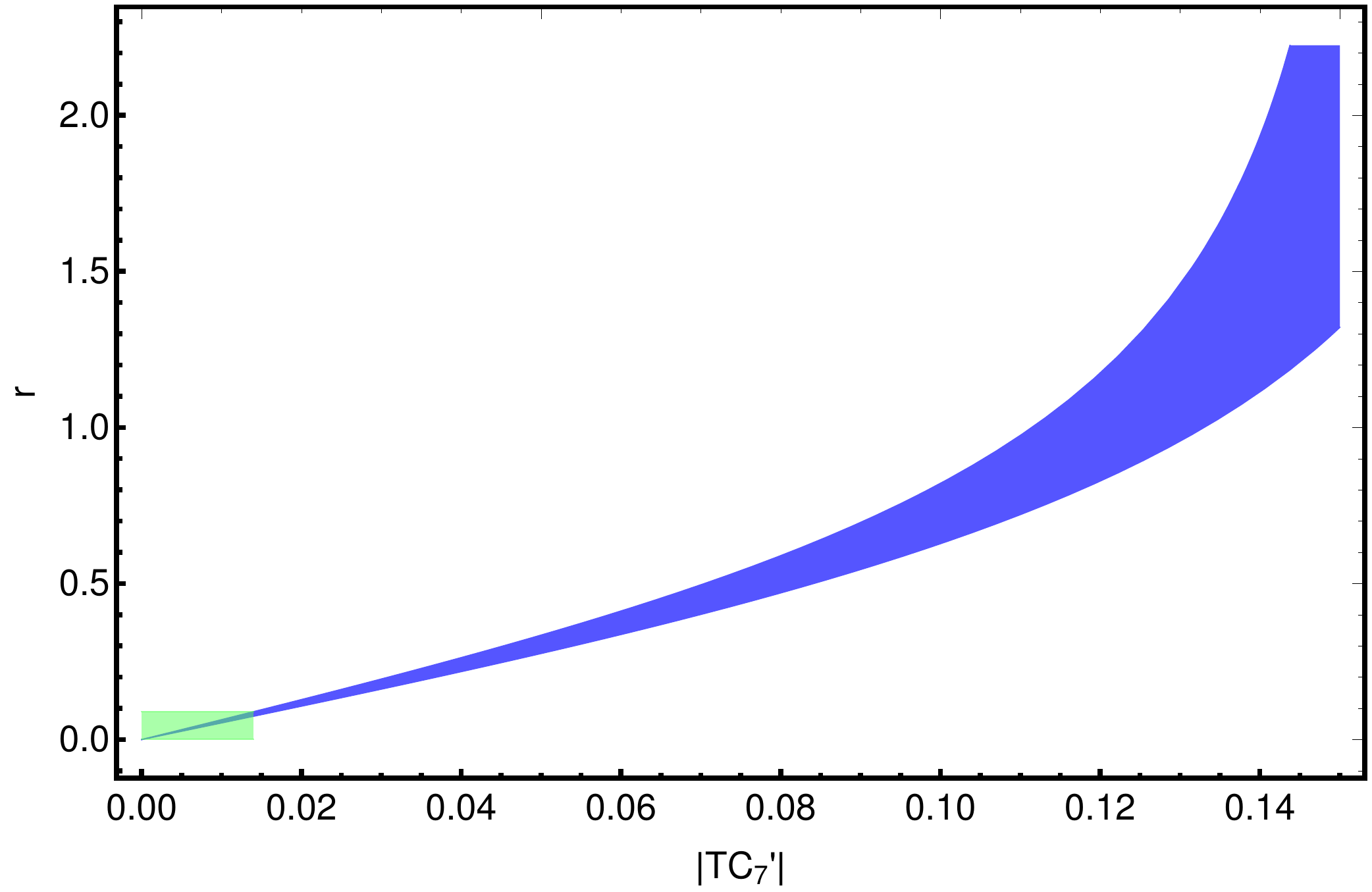}\qquad
 \includegraphics[width=0.45\textwidth]{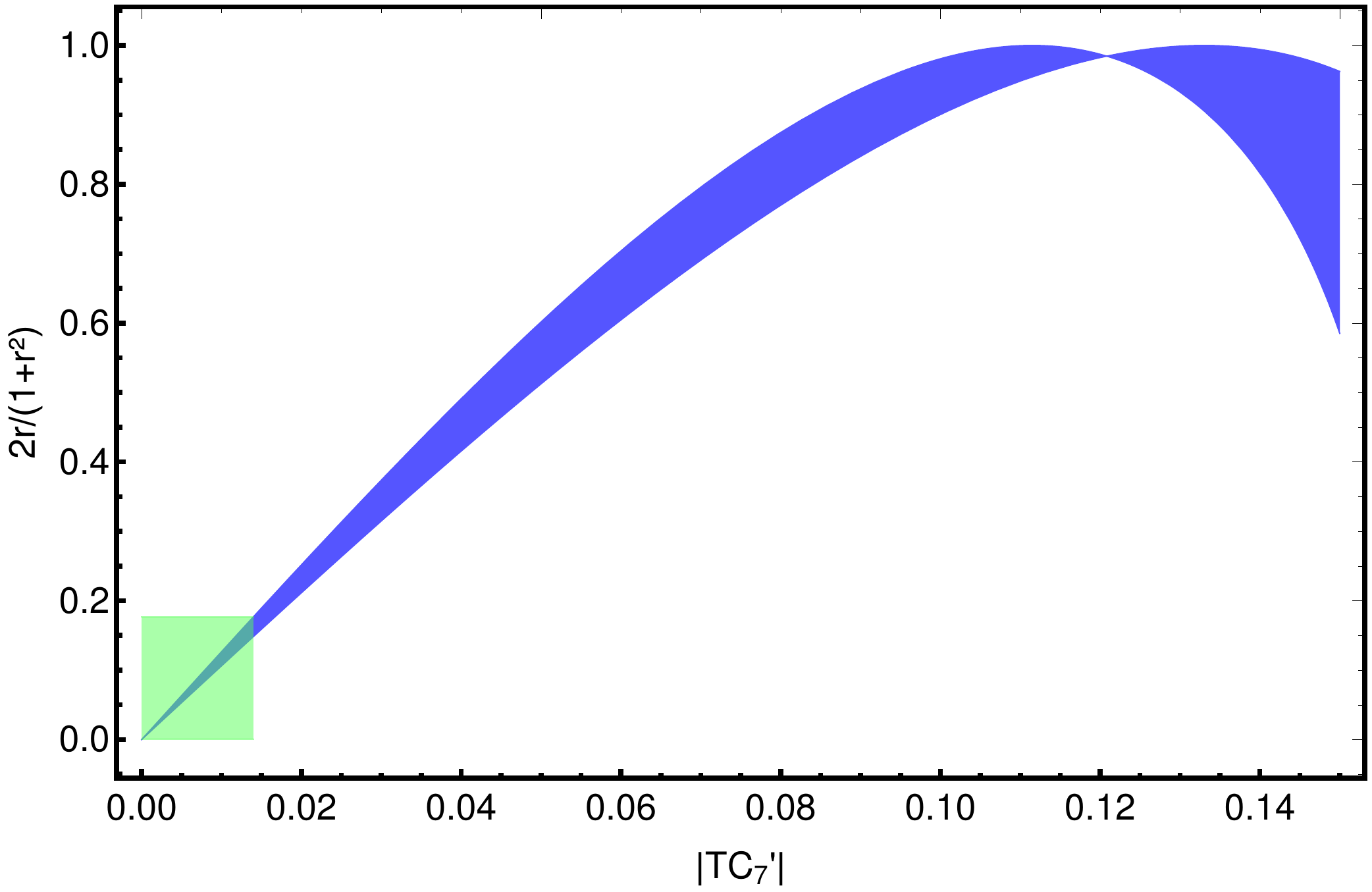}
 \caption{The polarization fraction $r$, eq.~(\ref{eq:r0zero}) and $2 r/(1+r^2)$, which drives $A^\Delta$, eq.~(\ref{sec:master}),  as a function of $| TC_7'|$ (blue shaded band) for the current data on ${\cal {B}}$ assuming $r_0\simeq0$. 
The range accessible by  leptoquark models is  indicated by the green box. 
 Model-independently, and in generic SUSY models, there is no upper limit on $r$.}
 \label{fig:r0zero}
\end{figure}

\item[c)] $C_7 \simeq 0$
\begin{align} \label{eq:rc}
r=\frac{ \sqrt{{\cal{B}} -|a|^2}}{ |a| }  \, . 
\end{align}
This allows to predict $r$   if one calculates  $a$. Using the results for the WA-amplitude in the heavy quark limit obtained in \cite{deBoer:2017que}
we find (for $\lambda_D\ge0.1\,\text{GeV}$) $r\ge2$. Such large values of $r$ are  consistent with the fact that the corresponding  SM prediction for the $D^0 \to \rho^0 \gamma$ branching ratio
${\cal{B}}=0.005 \cdot (0.1 \, \mbox{GeV}/\lambda_D)^2$ is significantly below the measured one given  in table~\ref{tab:br}.
Note that for the SM-dominated modes the agreement is much better,
${\cal{B}}_{\phi}=0.016  \cdot (0.1 \, \mbox{GeV}/\lambda_D)^2$ and ${\cal{B}}_{0}=0.16  \cdot (0.1 \, \mbox{GeV}/\lambda_D)^2$.
Here, $\lambda_D$ denotes a non-perturbative parameter expected to be of the order $\Lambda_{\rm QCD}$.

 \item[d)] $C_7^\prime \simeq 0$ 
\begin{align}
r= \frac{ r_0 |a|}{ \sqrt{ {\cal{B}}-r_0^2   |a|^2   } } \, . 
\end{align}
This will be useful once $r_0$ in addition to ${\cal{B}}_0$ is measured and, using eq.~(\ref{eq:Uspin}),  allows to illustrate viable ranges for $r$ in BSM scenarios.
\end{itemize}

Note, eq.~(\ref{eq:r0zero})  and  figure \ref{fig:r0zero}, and eq.~(\ref{eq:rc}) are independent of U-spin breaking. 

In the present situation where only branching fraction measurements are available, it is useful to define the ratio
\begin{align}
 R(\rho^0/\bar K^{*0})=\bigg | \frac{V_{cs}}{V_{cd}} \bigg |^2\frac{\cal{B}}{{\cal{B}}_{0}}=\frac{1+r^2}{1+r_0^2}\frac{{r'}^2}{r^2}  = 
  \frac{1+r^2}{1+r_0^2} |1+T C_7/a|^2\, , 
\end{align}
where $r'=|(a^{\prime}+T C_7')/a|$, and we used eq.~(\ref{eq:Uspin}).
The ratio $R$ equals one if only WA contributes,  that is, $r=r_0$ and $C_7=C_7^\prime=0$, irrespective of the size of the SM contributions. 
With BSM physics, the ratio can be larger or smaller than one.
Similar ratios have been mentioned in \cite{Bigi:1995em} as a test of the SM.

In $R$ only trivial U-spin breaking from phase space and CKM-elements has been accounted for.
We define in addition the ratio $\bar R$, in which also the leading dynamical one $\propto m_V f_V$ is covered.
\begin{align}
\bar  R(\rho^0/\bar K^{*0})=  \frac1{f^2}   \bigg | \frac{V_{cs}}{V_{cd}} \bigg |^2\frac{\cal{B}}{{\cal{B}}_{0}} \, , 
\end{align}
where, here, we use $f=m_\rho f_\rho/(m_{K^{*0}} f_{K^*}) \simeq 0.9$.

Using the data compiled in table~\ref{tab:br} and adding uncertainties in quadrature, we find
\begin{align} \label{eq:RroK}
 &R(\rho^0/\bar K^{*0})^\text{Belle}=1.14\pm0.21\,,&R(\rho^0/\bar K^{*0})^\text{BaBar}=1.61\pm0.33 \,,  \\
 \label{eq:RroKU}
 &\bar R(\rho^0/\bar K^{*0})^\text{Belle}=1.40 \pm0.26\,,&\bar R(\rho^0/\bar K^{*0})^\text{BaBar}=1.97 \pm0.40 \, .
\end{align}
Here, Belle and BaBar refers to the  respective measurement of the $D \to \bar K^{*0} \gamma$ branching ratio, which unfortunately, exhibit presently a 
significant experimental spread.
Inflating errors a la PDG  \cite{Patrignani:2016xqp}  due to the Belle/BaBar discrepancy, which exceeds one $\sigma$, we obtain for  the average $\bar R(\rho^0/\bar K^{*0})^\text{ave}=1.57\pm0.26$.

\begin{figure}
 \includegraphics[width=0.45\textwidth]{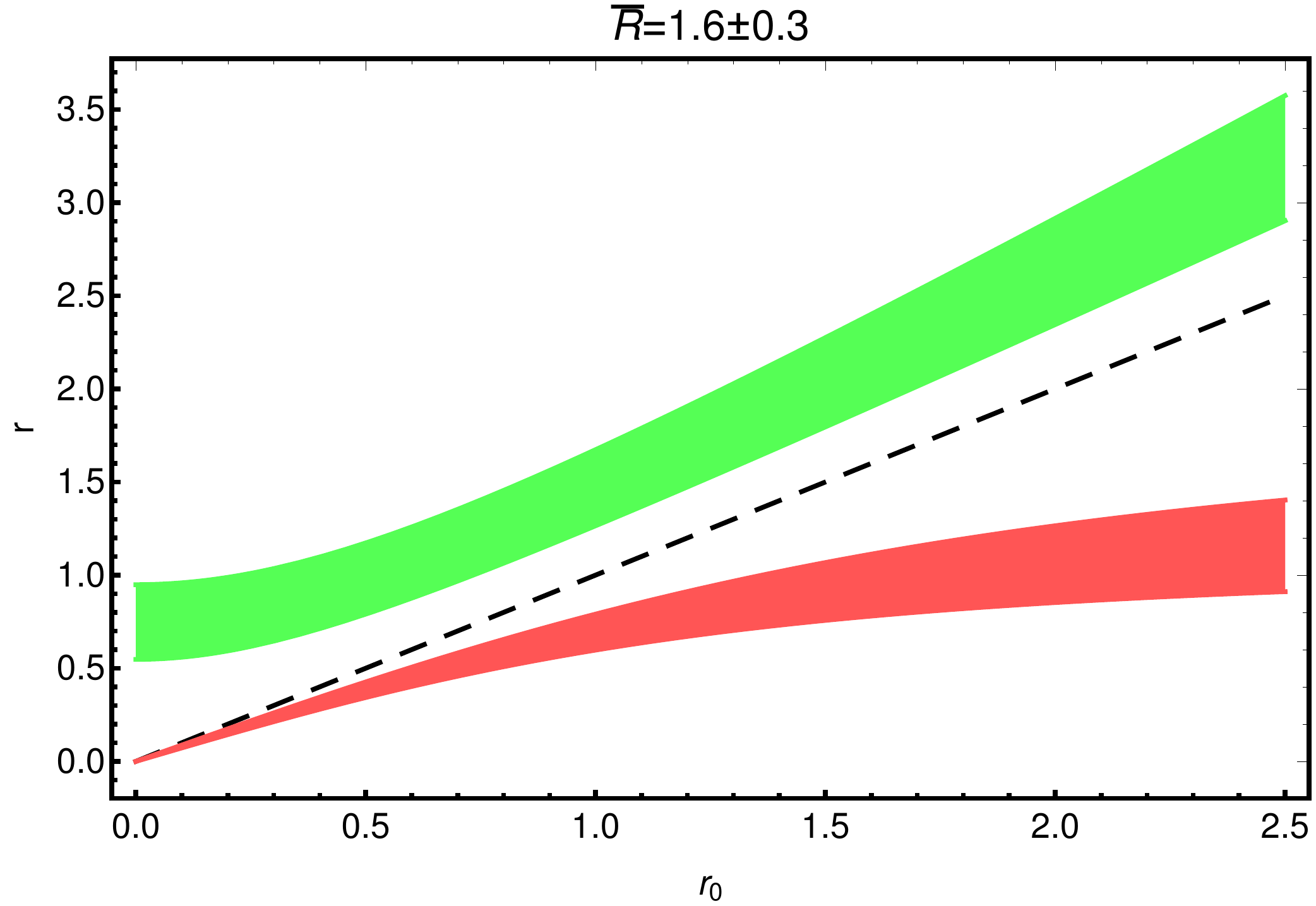}\qquad\includegraphics[width=0.45\textwidth]{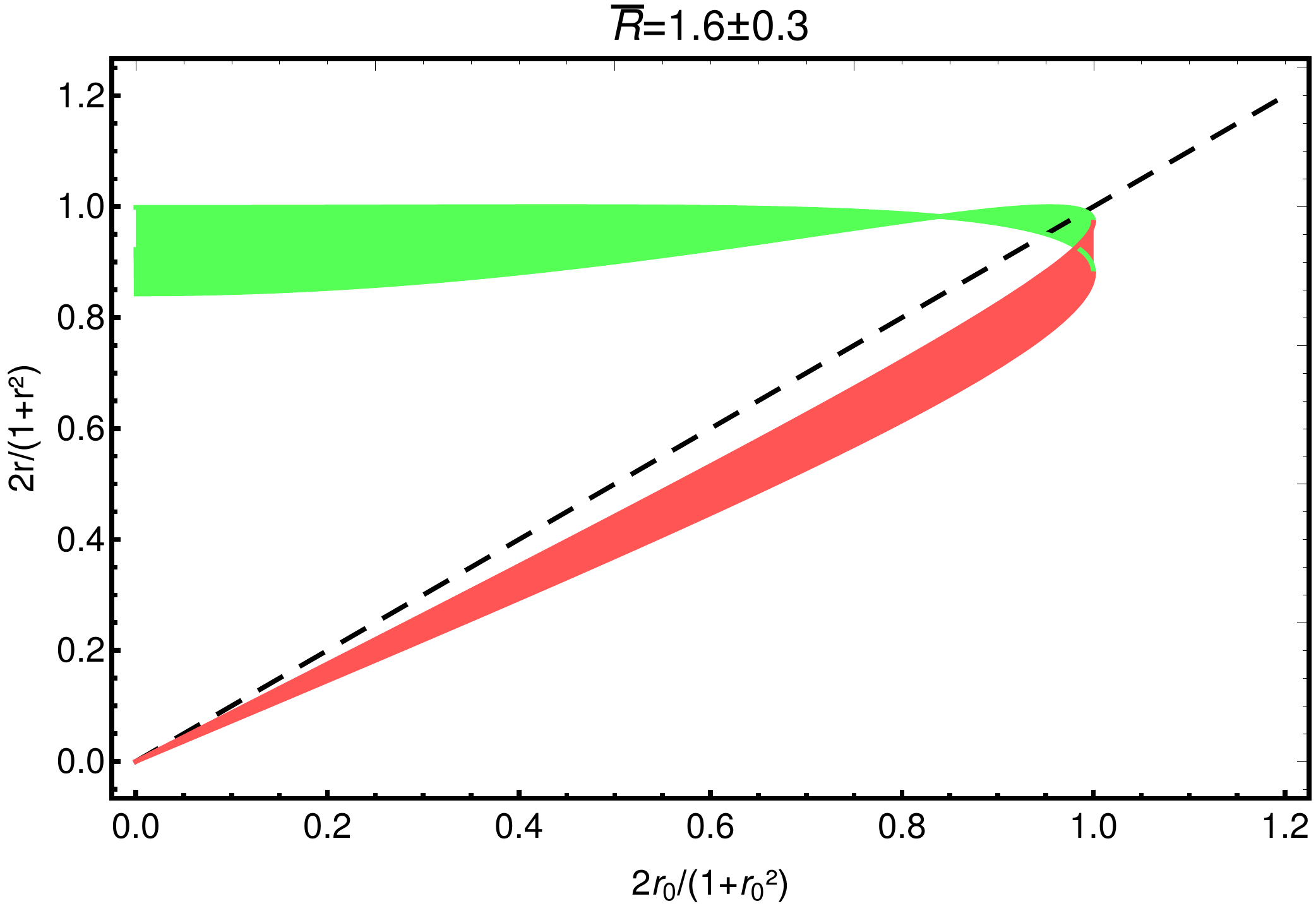}\\[2ex]
 \includegraphics[width=0.45\textwidth]{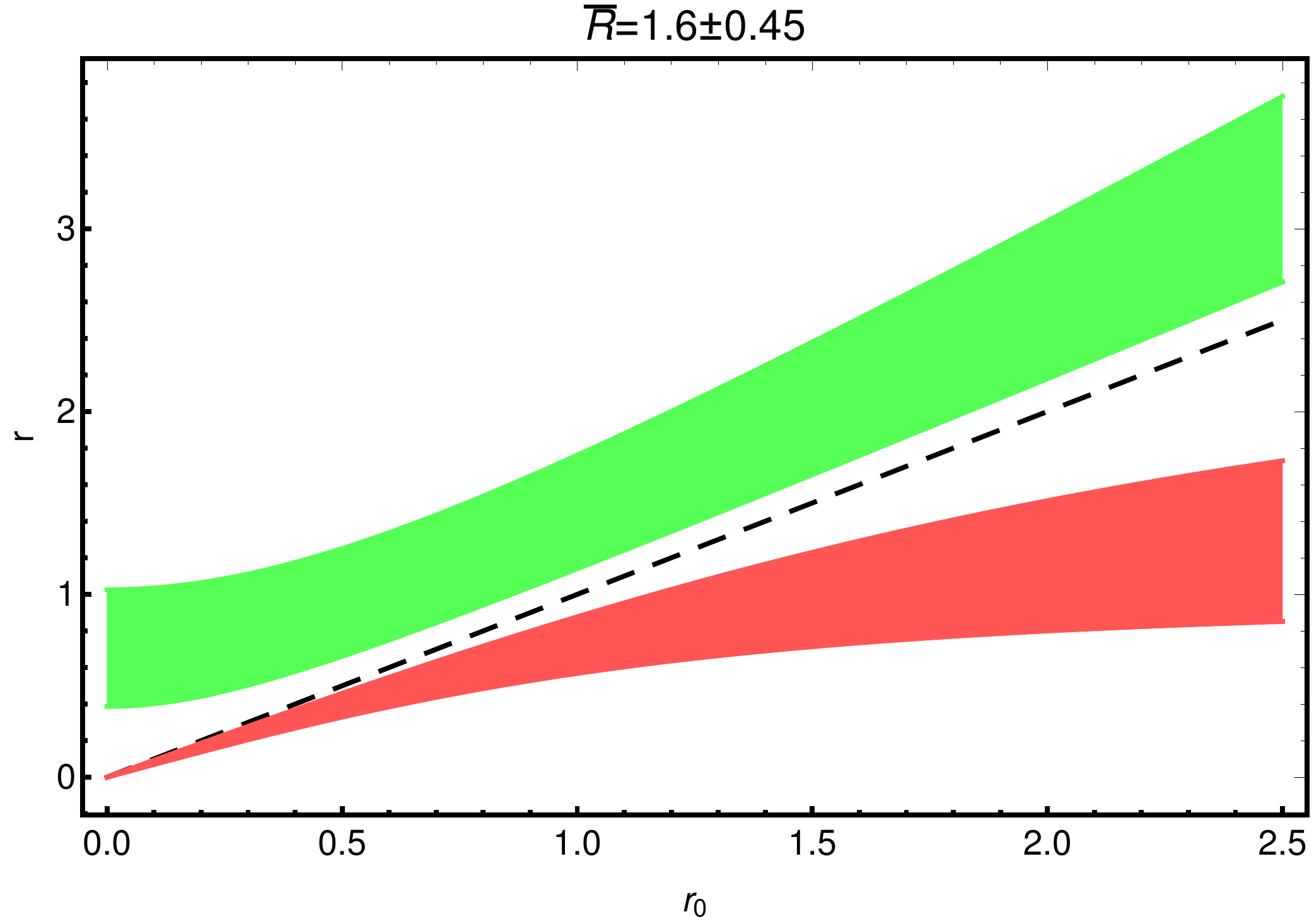}\qquad\includegraphics[width=0.45\textwidth]{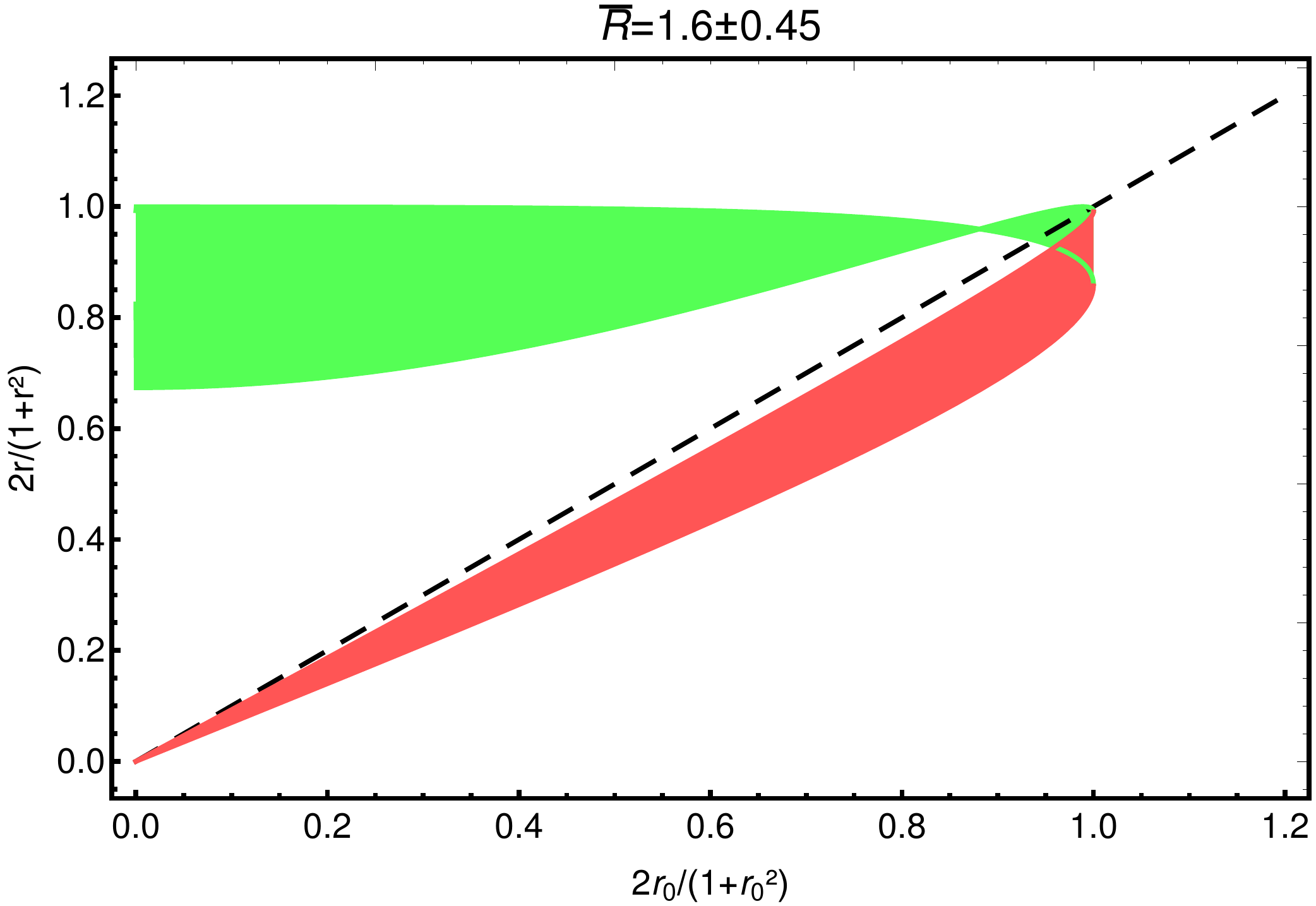}
 \caption{The polarization fraction $r$ as a function of $r_0$ (plots to the left) and $2r/(1+r^2)$ as a function of $2r_0/(1+r_0^2)$ (plots to the right), in the cases a) (SM case) $C_7,C_7'\simeq0$ (black, dashed curve),   c) $C_7\simeq0$  (green, upper band) and d) $C_7^\prime \simeq0$  (red, lower band).
 The upper (lower)  plots correspond to $\bar R=1.6\pm0.3$ ($\bar R=1.6\pm0.45$ from $50 \%$ inflated uncertainty).}
 \label{fig:Rdata}
\end{figure}

In figure~\ref{fig:Rdata} we show $r$ (plots to the left) for the cases a), c) and d) for $\bar R=1.6\pm0.3$ (upper row) and  $\bar R=1.6\pm0.45$ (lower row), illustrating the data's discriminative power.  
Case b) has already  been considered  in figure  \ref{fig:r0zero}. We learn that $r$ can be order one, and that it can be close to $r_0$, in which case discrimination from the SM is not possible.
In the plots to the right we show $2r/(1+r^2)$ as a function of $2r_0/(1+r_0^2)$,  which enter the observables $A^\Delta(D^0 \to \bar K^{*0} \gamma)$ and  $A^\Delta(D^0 \to \rho^0\gamma)$,
respectively, for a), c) and d). Within present data on  $\bar R$, scenario d) with no right-handed BSM physics (red band) cannot be sufficiently separated from the SM (black dashed curve), while
scenario c) with only $C_7^\prime$ present (green band), exhibits a  significant SM-deviation. 
For  $\bar R <1$ the green band corresponding to scenario c) would be below the SM curve while
the band corresponding to scenario d) would be above it. 
The lower plots in figure~\ref{fig:Rdata} correspond to a value of $\bar R$ with $50 \%$ larger uncertainty, mimicking larger U-spin corrections.
As the upper and lower plots are similar we learn that such effects do not change the picture qualitatively.

We can apply this strategy to probe for new physics in the decay $D \to \omega \gamma$, once its branching ratio and its polarization fraction become available.

\subsection{\texorpdfstring{$D^0 \to \phi  \gamma$}{D0tophigamma} \label{sec:fi}}

Due to the hybrid nature of the $\phi$,  one may ask whether
the $D \to \phi \gamma$ branching ratio is  consistent with the assumption of a predominantly WA-induced decay amplitude.
Corresponding ratios, 
\begin{align}
& R(\phi/\bar K^{*0})=\bigg | \frac{V_{ud}}{V_{us}} \bigg |^2\frac{\cal{B}_\phi}{{\cal{B}}_{0}} \, ,  && 
\bar R(\phi/\bar K^{*0})= \frac1{f^2} \bigg | \frac{V_{ud}}{V_{us}} \bigg |^2\frac{\cal{B}_\phi}{{\cal{B}}_{0}} \, , 
\end{align}
where ${\cal{B}}_\phi$ is the reduced branching ratio analogous to ${\cal{B}}$ for $D \to \rho^0 \gamma$, eq.~(\ref{eq:rhoB}),
and using $f =m_\phi f_\phi/(m_{K^{*0}} f_{K^*}) \simeq 1.3$,
are obtained as
\begin{align}
  &   R(\phi/\bar K^{*0})^\text{Belle}=1.46 \pm 0.15  \,  , &&  R(\phi/\bar K^{*0})^\text{BaBar}=2.10 \pm 0.37   \, ,  \\
  &   \bar R(\phi/\bar K^{*0})^\text{Belle}=0.87 \pm 0.09  \,  , &&  \bar R(\phi/\bar K^{*0})^\text{BaBar}=1.24 \pm 0.22   \,. 
\end{align}
Leading  U-spin breaking  makes  both  branching ratios of similar size. This is consistent with
 \begin{align}\bar R (\phi/\bar K^{*0}) =1+O(y) \, ,
\end{align}
which, together with eq.~(\ref{eq:rfi}), holds in the SM and beyond. 
 While this numerical agreement  could be accidental, it does give  a consistent picture between the predominantly SM-like modes and the size of
U-spin breaking in the range obtained within the lowest order heavy quark expansion \cite{deBoer:2017que},  sum rules  \cite{Khodjamirian:1995uc} and hybrid models  \cite{Fajfer:1997bh,Fajfer:1998dv}.
This is beneficial
as time-dependent analysis with $\bar K^{*0} \to K_{S,L} \pi^0$ is more difficult than with  $\phi \to K^+ K^-$.
We therefore suggest to use $r_\phi$, the polarization fraction of the photon in $D \to \phi \gamma$ decays, as a SM prediction for $r$.

We repeat the analysis previously performed with $\rho^0$ and $\bar K^{*0}$, eqs.~(\ref{eq:RroK}),(\ref{eq:RroKU}),  for the   $\rho^0$ and the $\phi$.
We obtain
\begin{align}
 &R(\rho^0/\phi)^\text{Belle}=0.78\pm0.17\,,&& R(\rho^0/\phi)^\text{BaBar}=0.77\pm0.21 \,  , \\
  &\bar R(\rho^0/\phi)^\text{Belle}=1.62\pm0.34\,,&& \bar R(\rho^0/\phi)^\text{BaBar}=1.59\pm0.43 \, , \label{eq:RtoPhiU}
\end{align}
and for  the average  $\bar R(\rho^0/\phi)^\text{ave}=1.61\pm0.27$. 
The good agreement seen in the data, between  eq.~(\ref{eq:RroKU}) and eq.~(\ref{eq:RtoPhiU}), supports that the $\phi$ indeed can be used as a standard candle as long as
effects of $O(y)$ can be neglected, and that the working assumption of the leading U-spin breaking is consistent with data.

\section{Summary \label{sec:con}}

Untagged, time-dependent analysis into CP eigenstates allows to extract the photon polarization in radiative charm decays.
Given a measurement of the photon's polarization fraction, its interpretation requires control of SM contributions to $D \to V \gamma$ decays.
We explore the possibility to obtain  the size of the SM background to wrong-chirality contributions from data and U-spin. While there are sizeable uncertainties related to this procedure,
there is presently no measurement available and  large room for BSM physics.

Specifically, we propose measurement of $A^\Delta$ in $D^0 \to \phi (\to K^+ K^-) \gamma$  decays to obtain the SM fraction, $r_\phi$.
While the $\phi$ is  not purely $s\bar s$ and therefore not purely WA-induced, the final state is advantageous over the one from the pure SM-mode,
$D^0 \to \bar K^{0*} (\to K_{S,L} \pi^0) \gamma$. If feasible, the latter should be studied experimentally as well.

If $r_\phi$, or $r_0$, the polarization fraction of  $D^0 \to \bar K^{0*} \gamma$ decays, is negligible, the photon polarization  and therefore $A^\Delta$ in $D^0 \to \rho^0 (\to \pi^+ \pi^-) \gamma$ becomes a null test of the SM. Possible ranges depending on the BSM model are illustrated in figure \ref{fig:r0zero}.
$A^\Delta$ in $D^0 \to \rho^0  \gamma$ decays can be ${\cal{O}}(1)$ in SUSY models, while leptoquark models give SM-like values.
The method works as well for $D^0 \to \omega \gamma$ decays, however, the branching ratios of the $\omega$ into suitable final states such as $\pi^+ \pi^-$ are small \cite{Patrignani:2016xqp}.

We further explored the correlation between the SM and BSM polarization fraction based on  ratios of branching fraction measurements, shown in figure \ref{fig:Rdata}.
Our study shows that U-spin breaking effects of nominal size are not qualitatively changing the picture.
In particular, we find that available branching ratios, table \ref{tab:br}, are consistent with U-spin hierarchies predicted by the heavy quark expansion, and other theory frameworks 
 \cite{Khodjamirian:1995uc,Fajfer:1997bh,Fajfer:1998dv}.
However, uncertainties are large, and more study  is needed to achieve a completer picture.
This includes the clarification of the discrepancy in the ${\cal{B}}(D^0 \to \bar K^{*0} \gamma)$ data.
A cleaner BSM interpretation would require better knowledge of the dipole form factor $T(0)$.

We point out that another way to probe the photon polarization in radiative charm decays  is provided by an up-down asymmetry (\ref{eq:updown}).
As in the time-dependent analysis  the SM value can be extracted from   a SM-like decay, here $D^0 \to \bar K_1^0 \gamma$, and then used together with U-spin for  a SM test in a BSM-sensitive mode,
$D_s \to K_1^+ \gamma$ decays.
Due to limited phase space the $K_1(1270)$ is more pronounced in charm relative to higher resonances than in $B$-decays.

The study of  the photon polarization complements BSM searches with CP asymmetries in $c \to u \gamma$ transitions.

\section*{Acknowledgements}
We are happy to thank Andrey Tayduganov for useful discussions.
This work has been supported  by the
DFG Research Unit FOR 1873 ``Quark Flavour Physics and Effective Field Theories''
and by the BMBF under contract no. 05H15VKKB1.

\appendix

\section{Up-down asymmetry}

The differential  distribution of $D\to R\gamma\to P_1P_2P_3\gamma$  decays via a $J^P=1^+$ resonance $R$ can be written as \cite{Gronau:2001ng,Gronau:2002rz,Kou:2010kn,Gronau:2017kyq}
\begin{align}
 \frac{\mathrm d\Gamma}{\mathrm ds_{13}\,\mathrm ds_{23}\,\mathrm d\cos\theta}\propto|\vec J|^2(1+\cos^2\theta)+\lambda_\gamma2\,\mathrm{Im}[\vec n\cdot(\vec J\times \vec J^*)]\cos\theta\,,
\end{align}
where $\lambda_\gamma=(|\mathcal A_R|^2-|\mathcal A_L|^2)/(|\mathcal A_R|^2+|\mathcal A_L|^2)$ denotes the photon polarization parameter, 
$s_{ij}=(p_i+p_j)^2$ with the four momenta $p_i$ of the mesons $P_i$ and $\theta$ is the angle between the normal $\hat n=((\vec p_1\times\vec p_2)/|(\vec p_1\times\vec p_2)|)$ and the direction opposite to the photon in the rest frame of $R$.
The integrated up-down asymmetry reads
\begin{align}
 A_{up-down}&=\left(\int_0^1\frac{\mathrm d\Gamma}{\mathrm d\cos\tilde\theta}\mathrm d\cos\tilde\theta-\int_{-1}^0\frac{\mathrm d\Gamma}{\mathrm d\cos\tilde\theta}\mathrm d\cos\tilde\theta\right)\bigg/\int_{-1}^1\frac{\mathrm d\Gamma}{\mathrm d\cos\tilde\theta}\mathrm d\cos\tilde\theta\nonumber\\
 &=\frac34\frac{\left<\mathrm{Im}[\hat n\cdot(\vec J\times\vec J^*)]\,\mathrm{sgn}[s_{13}-s_{23}]\right>}{\left<|\vec J|^2\right>}\lambda_\gamma\,,
 \label{eq:updown}
\end{align}
where $\cos\theta=\mathrm{sgn}[s_{12}-s_{23}]\cos\tilde\theta$ and the $\left<  \, .. \, \right>$-brackets denote integration over $s_{13}$ and $s_{23}$.
Here, $\vec J$ is defined by the decay amplitude $\mathcal A(R\to P_1P_2P_3)=\epsilon^\mu J_\mu$ with the polarization vector $\epsilon$ of $R$.
Formulas for $J$ including contributions from resonances with different spin and parity, {\it e.g.,}~the $K(1400)$-family and their interference effects can be extracted from \cite{Gronau:2001ng,Gronau:2002rz,Kou:2010kn,Gronau:2017kyq}.
Decay chains involving a kaon resonance are collected in table \ref{tab:decay_chains}. 
Decays of $D^+$ and $D_s$ are self-tagging.

\begin{table}
 \centering
 \begin{tabular}{lccc}
  \toprule
  decay chain                                           &  color & CKM & type  \\
  \midrule
  $D^0\to\bar K_1^0\gamma\to\bar K \pi\pi\gamma$  &  $1/N_C$ &  CF &  SM \\
  $D^0\to K_1^0\gamma\to K\pi\pi\gamma$               &  $1/N_C$ & DCS & SM \\
 $D^+\to K_1^+\gamma\to K\pi\pi\gamma$          &  1         & DCS & SM \\
$D_s\to K_1^+\gamma\to K\pi\pi\gamma$      &  1          & SCS & FCNC\\
  \bottomrule
 \end{tabular}
 \caption{Decay chains involving a kaon resonance decaying via $K_1^+\to(K^{*+}\pi^0,K^{*0}\pi^+,\rho^+K^0)\to K^0\pi^+\pi^0$, $K_1^+\to(K^{*0}\pi^+,\rho^0K^+)\to K^+\pi^+\pi^-$, $K_1^0\to(K^{*+}\pi^-,K^{*0}\pi^0,\rho^-K^+)\to K^+\pi^-\pi^0$ and $K_1^0\to(K^{*+}\pi^-,\rho^0K^0)\to K^0\pi^+\pi^-$. 
 Here, CF, SCS and DCS denote Cabibbo-favored, singly Cabibbo-suppressed and doubly Cabibbo-suppressed WA-amplitudes, respectively. While 'SM' indicates a WA-induced decay, 'FCNC' indicates the presence of
 $c \to u$ contributions.}
  \label{tab:decay_chains}
\end{table}

\end{document}